\documentclass[10pt,leqno]{amsart}
\usepackage{graphicx}
\usepackage{indentfirst,csquotes}

\usepackage{amssymb,amsthm,amsmath}
\usepackage{xcolor,paralist,hyperref,fancyhdr,etoolbox}
\usepackage[numbers,sort&compress]{natbib} 

\usepackage{titlesec}
\titleformat{\section}{\bfseries\Large}{\thesection.}{1em}{}
\titleformat{\subsection}{\bfseries\large}{\thesubsection.}{1em}{}
\titlespacing*{\subsection}{0pt}{3.5ex plus 1ex minus .2ex}{1.5ex plus .5ex minus .2ex}

\hypersetup{ colorlinks=true, linkcolor=black, filecolor=black, urlcolor=black }

\begin{document}

\title{Introduction to Quantum Ophthalmology}

\author{
Mukhit Kulmaganbetov$^{1,2,3,*}$\and
Dmitry Pushin$^{4,5,6,7}$\and
Taranjit Singh$^{1,2}$\and
Pinki Chahal$^8$\and
David Cory$^{4,9}$\and
Iman Salehi$^{1,4,6}$\and
Andrew Silva$^{6,10}$\and
Ben Thompson$^{1,6}$\and
Dusan Sarenac$^{6,7,8}$}

\footnotetext[1]{Quantum Optics Lab, Centre for Eye and Vision Research, Hong Kong, PR China}
\footnotetext[2]{Entoptica Limited, Hong Kong, PR China}
\footnotetext[3]{Department of Molecular Biology and Medical Genetics, Kazakh National Medical University, Almaty, Kazakhstan}
\footnotetext[4]{Institute for Quantum Computing, University of Waterloo, Waterloo, Canada}
\footnotetext[5]{Department of Physics and Astronomy, University of Waterloo, Waterloo, Canada}
\footnotetext[6]{School of Optometry and Vision Science, University of Waterloo, Waterloo, Canada}
\footnotetext[7]{Incoherent Vision Inc., Wellesley, Canada}
\footnotetext[8]{Department of Physics, University at Buffalo, State University of New York, Buffalo, USA}
\footnotetext[9]{Department of Chemistry, University of Waterloo, Waterloo, Canada}
\footnotetext[10]{Department of Psychology, Idaho State University, Pocatello, USA}

\maketitle

\let\thefootnote\relax

\begin{abstract}
Quantum technologies are rapidly advancing across multiple research domains, with a growing impact on biomedical imaging and sensing. We examine their emerging role in ophthalmology through four complementary directions: photon-limited retinal imaging, correlation-based imaging, nanoscale optical probes, and quantum-limited visual perception. Advances in optical coherence tomography and single-photon detection enable imaging under strict photon budget constraints, reducing phototoxicity while preserving image quality. Correlation-based approaches, including ghost imaging, offer alternative strategies for image formation in low-light and scattering environments, although practical implementation remains limited by detection efficiency and acquisition time. In parallel, nanoscale optical platforms such as quantum dots provide tunable and photostable probes for enhanced contrast and targeted delivery, with ongoing challenges related to biocompatibility and clinical translation. Finally, experiments at the single-photon level and with structured light fields demonstrate how the visual system itself operates near physical detection limits and can be probed using controlled optical states. While many of these approaches remain at an early stage, they collectively illustrate how quantum and quantum-inspired methods may augment current ophthalmic imaging and diagnostic technologies while providing new tools for studying visual function under well-defined physical constraints.
\end{abstract}

\bigskip

\noindent \textbf{Keywords:} Quantum Ophthalmology, Quantum Imaging, Optical Coherence Tomography, Ghost Imaging, Quantum Dots, Visual Perception.

\section{Introduction}
Quantum technologies refer to approaches that exploit fundamental properties of quantum systems, such as superposition, coherence, entanglement, and quantization, to enable new capabilities in computing, communication, and sensing \cite{upadhyay2025, chinnappan2025, khan2024, engelsberger2023, yago2024, gilaberte2019}. Over the past decades, these concepts have transitioned from foundational physics into practical tools, driving advances in communication security \cite{Narottama2023,Fernandez2020}, metrology \cite{Montenegro2025,Schnabel2010}, sensing \cite{Aslam2023,Degen2017}, and computing \cite{Feynman1982, Deutsch1985, Gill2022}. In the biomedical context, quantum principles underpin established techniques such as functional magnetic resonance imaging (fMRI) \cite{grover2015magnetic, lauterbur1973image, zia2019nuclear}, which relies on nuclear spin dynamics for non-invasive measurement of biological structure, and electron microscopy \cite{ruska1987development, dwyer2023quantum, akhtar2025advancing, klang2013electron}, where the wave nature of electrons enables nanometer-scale imaging that supports structural biology and drug discovery. Furthermore, magnetoencephalography (MEG), which uses a superconducting quantum interference device (SQUID), serves as a prominent example of quantum-enabled medical sensing \cite{bonavolonta2025, schofield2022, ren2025}. These examples illustrate how quantum physics has long contributed to high-resolution imaging and precision measurement in medicine, providing a foundation for emerging quantum-enhanced approaches. Throughout this review, we distinguish between nonclassical-light-based approaches, quantum-enabled detection technologies, and quantum-inspired optical methods, which play complementary but conceptually distinct roles.

In ophthalmology and vision science, quantum effects are relevant both at the level of light detection and in the interaction of light with retinal tissue. The human visual system operates near the physical limit of sensitivity, with rod photoreceptors capable of responding to individual photons under appropriate conditions \cite{rieke1998single, phan2014interaction, gassab2024conditions, tinsley2016direct}. This extreme sensitivity may provide a natural platform for studying light detection at the quantum limit and has motivated the use of controlled single-photon sources to probe visual perception. In parallel, nonseparable optical fields carrying orbital angular momentum have enabled new experiments exploring how the eye responds to structured light \cite{pushin2025characterizing, pushin2023structured, pushin2024psychophysical}.

At the molecular level, phototransduction is initiated by photoisomerisation of opsin molecules \cite{blanco2024comparative}. The efficiency of this process is characterized by its quantum yield \cite{stranius2017determining}, which nature optimizes via steric constraints in proteins like rhodopsin \cite{shieh1997steric}. Valentini et al. (2017) further demonstrated that tensile forces on retinal models substantially increase the trans-to-cis photoisomerization yield, providing a rationale for how extension forces facilitate the formation of the compressed cis-isomer \cite{valentini2017optomechanical}. This is a process that can be understood through quantum transitions in retinal chromophores. More broadly, concepts from quantum biology, such as coherence in energy transfer and tunnelling in biochemical reactions \cite{engel2007evidence,klinman2013hydrogen}, have been discussed in the context of biological light-matter interactions. While the direct role of such effects in visual function remains an open question, these perspectives have stimulated new approaches to optical probing and sensing in biological systems.

Biomedical imaging in ophthalmology has been transformed by techniques such as optical coherence tomography (OCT) \cite{fujimoto2000optical}, confocal microscopy \cite{chiang2023invivo}, and fluorescence imaging \cite{dysli2017fluorescence}, which enable non-invasive visualization of retinal and other ocular structures with high spatial resolution. Despite their success, these methods face practical constraints arising from the properties of classical light sources and detectors, including limited sensitivity at low photon flux, trade-offs between resolution and illumination dose, and the risk of photodamage in light-sensitive tissues \cite{morgan2008light,zhong2016fundus}. These considerations are particularly important in ophthalmic imaging, where exposure levels must remain within strict safety limits.

Quantum and quantum-inspired imaging approaches have been proposed as a means to address some of these challenges. Techniques based on nonclassical states of light, including entangled and correlated photon sources and squeezed states, offer new strategies for enhancing measurement sensitivity and improving signal-to-noise performance under photon-limited conditions \cite{defienne2024advances, schwartz2013superresolution, zhang2024quantum, samimi2025quantum, aarav2025imaging}. In parallel, advances in quantum optics hardware, such as bright single-photon sources and high-efficiency detectors \cite{aharonovich2016solid, esmann2024solid, hadfield2023single, dao2025single}, are enabling experimental implementations that approach regimes difficult to access with conventional systems. While the extent of quantum advantage in practical biomedical imaging remains an active area of investigation, these developments are beginning to bridge the gap between fundamental quantum optics and applied imaging technologies.

While practical challenges remain, including the development of robust quantum light sources and detectors compatible with clinical environments, rapid progress in photonics and computational methods is accelerating the translation of these approaches into ophthalmology. In this review, we organize recent developments into four complementary directions: photon-limited retinal imaging, including advances in OCT and single-photon detection; correlation-based imaging approaches, such as ghost imaging, that enable image formation under low-light or nonstandard geometries; nanoscale optical probes, including quantum dots, for enhanced contrast and targeted delivery; and the use of controlled optical fields to probe visual perception at the limits of sensitivity. Across these areas, we emphasise how quantum and quantum-inspired methods extend existing imaging capabilities, provide new tools for probing visual function, and highlight the technical challenges that must be addressed to enable clinical adoption.

\section{Retinal imaging and phototoxicity}
\subsection{Current status}
OCT is a high-resolution imaging technique that uses low-coherence interferometry to generate three-dimensional images of both anterior and posterior eye structures \cite{drexler1999in}. It has become a central diagnostic tool in ophthalmology due to advances in acquisition speed and improvements in axial and lateral resolution \cite{fujimoto2000optical, Kulmaganbetov2025}. Despite its success, OCT performance is ultimately constrained by the allowable photon flux: safety standards limit retinal exposure, restricting the signal-to-noise ratio and image quality that can be achieved in practice \cite{fujimoto2000optical, drexler1999in}. Increasing illumination power to improve sensitivity is therefore not generally feasible due to the risk of retinal phototoxicity.

\subsection{Single-photon detection approaches}
One route to improving OCT under photon-limited conditions is through the use of highly sensitive detection schemes. Kolenderska et al. (2020) demonstrated an OCT system incorporating a superconducting single-photon detector (SSPD) and field-programmable gate arrays (FPGA)-based time-stamping electronics, combined with dispersive Fourier transformation \cite{kolenderska2020quantum}. In this setup (Figure 1A), a pulsed laser attenuated to a low mean photon number per pulse is coupled into a fibre and propagates through a Linnik–Michelson interferometer. The input pulse is split at a beam splitter into object and reference arms, where the object arm interacts with the sample to acquire an additional phase, while the reference arm is reflected from a mirror. The two paths overlap at the beam splitter, and the output is coupled to a single-mode fibre spool using a fibre coupler. The time-resolving SSPD, together with the long dispersive fibre spool, works as a spectrometer. Time reference is provided by a photodiode signal from the light source, and the data are collected using FPGA time-stamping electronics.  This approach enables high-quality imaging at incident power levels on the order of 10 pW, achieving image quality comparable to conventional systems while significantly reducing light exposure. Such detector-level advances, which include semiconductor nanowire avalanche photodiodes that leverage internal carrier multiplication to detect single excitons \cite{bulgarini2012avalanche}, directly address the photon budget limitation and are particularly relevant for ophthalmic imaging, where minimizing phototoxicity is critical.

\begin{figure}[htbp]
 \centering
        \includegraphics[width=0.7\textwidth]{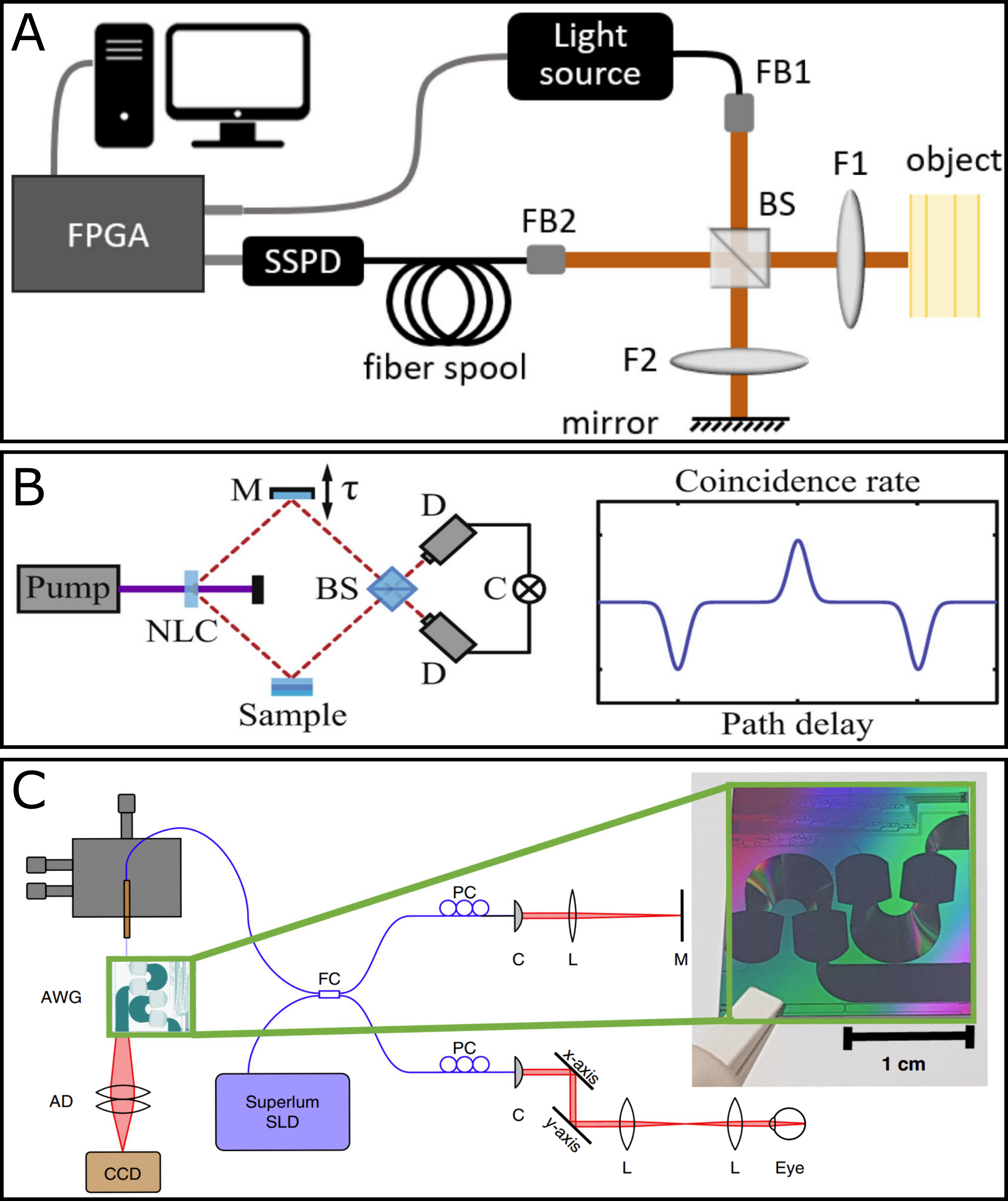}
 \caption{Advanced optical coherence tomography (OCT) modalities for addressing retinal phototoxicity. A. Experimental setup for quantum-inspired OCT. A pulsed laser attenuated to a low mean photon number per pulse is coupled into a Linnik–Michelson interferometer. Light in the object arm interacts with the sample, while the reference arm reflects from a mirror. The combined output is coupled into a dispersive fibre spool. A superconducting single-photon detector (SSPD) and the dispersive fibre function as a spectrometer, with data acquisition via field-programmable gate arrays (FPGA) time-stamping electronics. Adapted with permission from \cite{kolenderska2020quantum} © Optical Society of America. B. Quantum OCT (Q-OCT) using spontaneous parametric down-conversion (SPDC) photon pairs. (left) Schematic of the setup where signal photons reflected from the sample interfere with idler photons carrying a temporal delay at a beam splitter (BS). (right) Example of a Q-OCT interferogram for a two-layer sample, where the separation of the dips is proportional to the sample optical thickness. The central peak is due to the cross-correlation between the reflection of the probe photon by both layers. Adapted from \cite{graciano2019interference}. C. Scheme of the spectral domain optical coherence tomography (SD-OCT) on-chip setup. Superluminescent diode (SLD) feeds broadband light to a fibre coupler. Light on the eye (e.g., 830$\mu$W or 480$\mu$W) interferes with the reference light and is coupled into the on-chip arrayed waveguide grating (AWG). Projection optics project the light from the photonic integrated circuit (PIC) end facet onto a charge-coupled device (CCD) camera. Adapted from \cite{rank2022miniaturizing}.}
\label{fig1}
\end{figure}

\subsection{Quantum OCT}
Beyond detector-level improvements, uniquely quantum approaches have also been explored (Figure 1B). Quantum optical coherence tomography (Q-OCT) employs entangled photon pairs, typically generated via spontaneous parametric down-conversion (SPDC), to perform correlation-based interferometric measurements \cite{teich2012variations, kulkarni2022classical, mohan2009ultrabroadband}. Graciano et al. (2019) demonstrated a Q-OCT scheme where signal photons reflected from the sample interfere at the beam splitter with idler photons carrying a temporal delay \cite{graciano2019interference}. The resulting interferogram for a two-layer sample exhibits dips separated by a distance proportional to the sample optical thickness, with a central peak arising from the cross-correlation between the reflection of the probe photon by both layers.

Beyond Q-OCT, uniquely quantum approaches leverage photon correlations to achieve pixel super-resolution beyond detector limits \cite{defienne2022pixel} and offer enhanced signal-to-noise ratios for imaging under low-light conditions \cite{vernekar2024quantum}, while fundamental experiments continue to probe the nature of two-photon interference \cite{kaur2020quantum}. While Q-OCT has demonstrated proof-of-principle advantages in controlled settings, its practical benefits for biomedical imaging remain under active investigation, particularly in terms of achievable resolution, acquisition speed, and system complexity.

\subsection{Miniaturization and integration}
Translating these approaches into clinical practice requires compact, stable, and scalable implementations. Recent work in integrated photonics has demonstrated the feasibility of miniaturized OCT systems based on silicon photonic platforms \cite{rank2022miniaturizing, schneider2016optical}. For instance, Rank et al. (2021) implemented an spectral domain OCT on-chip setup (Figure 1C) capable of in vivo retinal imaging with high sensitivity, utilizing power levels on the eye of approximately 480 $\mu$W or 830 $\mu$W depending on the configuration \cite{rank2022miniaturizing}. These systems benefit from compatibility with complementary metal-oxide-semiconductor (CMOS) fabrication, enabling scalable integration of optical components, such as light sources, waveguides, modulators, and switches, and electronic control \cite{rank2022miniaturizing, gupta2022silicon}. For quantum and single-photon-based approaches, photonic integration offers a pathway toward improved stability, reduced system complexity, and eventual clinical deployment. Optimizing the photonic environment of these integrated emitters is crucial for maximizing signal; recent cavity designs utilizing quasi-bound states in the continuum have demonstrated enhanced extraction efficiency and directional emission for nanowire quantum dots \cite{gangopadhyay2026broadband}. Although challenges remain in integrating efficient sources and detectors within practical imaging systems, hybrid techniques that deterministically integrate preselected nanowire quantum dots onto silicon photonic circuits have demonstrated high coupling efficiencies and maintained single-photon purity \cite{esmaeilzadeh2016deterministic}, bridging the gap between quantum light sources and scalable chip-based architectures.

\section{Volumetric and low-light tissue visualization}
\subsection{Current status}
In ophthalmic imaging, resolving the three-dimensional structure of the retina typically relies on mechanical scanning \cite{cao2016design, yasuno2005three}, which can limit acquisition speed and increase total light exposure. Imaging in scattering environments, such as ocular tissue, further degrades signal quality and contrast. These constraints are particularly significant in ophthalmology, where illumination levels must remain low to avoid phototoxicity, making it challenging to obtain volumetric information under photon-limited conditions \cite{ha2012low, franco2024phototoxicity}.

Beyond ghost imaging, computational frameworks utilizing single-photon avalanche diodes (SPADs) have been developed to reconstruct hidden or obscured targets by correlating transient time-of-flight data. Recent algorithms, such as the confocal complemented signal-object collaborative regularization (CC-SOCR), have demonstrated the ability to reconstruct high-fidelity images from arbitrary, sparse illumination patterns \cite{Liu2023Non-line-of-sight}, highlighting the potential of advanced data processing to overcome photon budget limitations.

\subsection{Quantum ghost imaging}
Ghost imaging offers an alternative imaging paradigm based on correlation measurements rather than direct spatial detection. In quantum implementations, spatial information is reconstructed from correlations between photon pairs (Figure 2A), allowing image formation even when the detected photons do not directly interact with the spatial structure of the sample \cite{shapiro2012, mavian2025, padgett2017}.

\begin{figure}[htbp]
 \centering
        \includegraphics[width=0.65\textwidth]{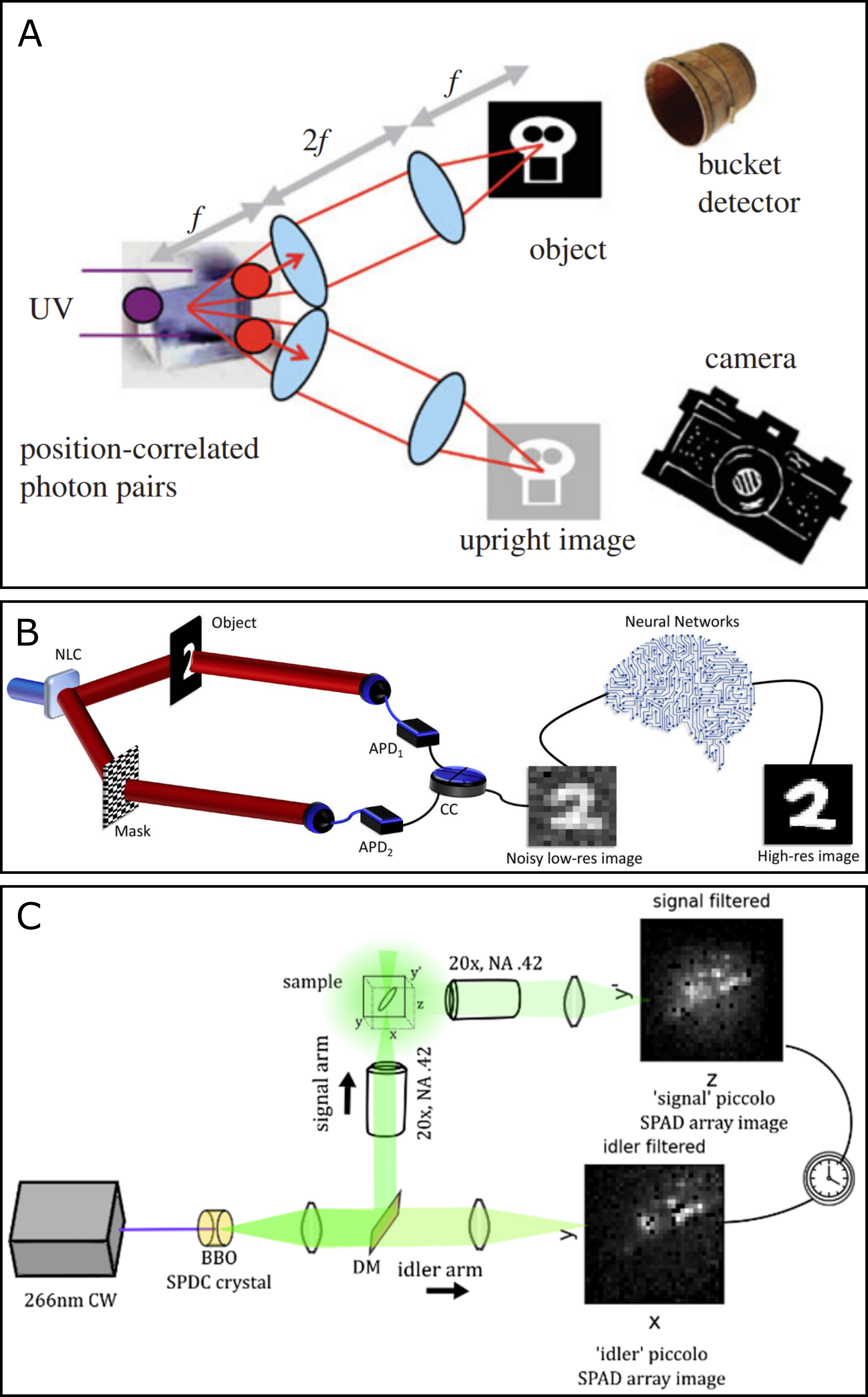}
 \caption{Advances in quantum and computational ghost imaging for volumetric and low-light visualization. A. Conceptual comparison of ghost imaging systems. Images produced by a ghost imaging system based on spontaneous parametric down-conversion (SPDC) are equivalent to those produced by a classical imaging system. However, the ghost imaging system operates with a different time sequence of events. Adapted with permission from \cite{padgett2017}. B. All-digital ghost imaging setup for super-resolution. Entangled photons are spatially separated along two arms. One photon interacts with the object and is collected by a bucket detector, while the other is collected by a spatially resolving detector comprising a patterned mask and a bucket detector. Detectors are connected to a coincidence-counting device, and neural networks process the reconstructed low-resolution image for denoising and super-resolution. Adapted from \cite{moodley2022super}. C. Schematic of a three-dimensional ghost imaging microscope. Signal and idler photons are separated into distinct optical paths. Idler photons travel directly to a single-photon avalanche diode (SPAD) array. The signal path utilizes two microscope objectives placed orthogonal to each other; the first focuses light onto the sample, and the second captures scattered light measured by a second SPAD array. Adapted with permission from \cite{eshun2025} © Optical Society of America}
\label{fig2}
\end{figure}

Furthermore, computational approaches have enabled super-resolution capabilities; Moodley and Forbes (2022) demonstrated an all-digital ghost imaging setup where entangled photons are separated into two arms: (1) one interacting with an object (bucket detector) and (2) the other passing through a patterned mask (spatially resolving detector). The reconstructed low-resolution image is subsequently processed using neural networks (Figure 2B) for denoising and super-resolution \cite{moodley2022super}. Recent developments have focused on improving efficiency and acquisition speed \cite{chang2025improving, wang2017highspeed} and addressing moving targets \cite{li2015ghost}, moving ghost imaging closer to practical applications in low-light and scattering environments \cite{ryan2024infrared, zhang2025scattering}. However, correlation-based imaging remains limited by detection efficiency and measurement time, which currently constrain its applicability in real-time biomedical imaging.

\subsection{Three-dimensional ghost imaging}
Extending ghost imaging to volumetric reconstruction, Eshun et al. (2025) demonstrated a three-dimensional ghost imaging microscope (Figure 2C) based on time-correlated single-photon detection \cite{eshun2025}. In this configuration, signal and idler photons are separated into different optical paths. The idler photons travel directly to a SPAD array. The signal path incorporates two microscope objectives placed orthogonal to each other: the first objective lens focuses the light onto the sample, and the second objective lens captures the scattered light, which is measured at the second SPAD array. By recording the arrival times of correlated photons in separate detection paths, the system reconstructs orthogonal projections of the sample and combines them to recover volumetric information without mechanical scanning. This approach enables three-dimensional imaging at low light levels and illustrates a pathway toward non-scanning volumetric reconstruction.

While current implementations remain proof-of-principle, the ability to access depth information through time-correlated measurements is particularly relevant for ophthalmology, where minimizing light exposure is critical. These methods may provide new routes to imaging delicate retinal structures under conditions where conventional scanning techniques are limited.

\subsection{Computational and human vision integration}
Related developments in computational ghost imaging highlight connections between correlation-based imaging and temporal integration in human vision. Experiments have shown that the human visual system can integrate structured illumination patterns over time to form recognizable images without explicit computational reconstruction \cite{boccolini2019ghost, wang2020all}. These results demonstrate that image formation can arise from temporal accumulation of correlated signals, offering insight into both computational imaging strategies and the processing of structured light by the visual system.

\section{Quantum dots for imaging and targeted delivery}
\subsection{Current status}
In ophthalmic practice, diagnostic imaging often relies on fluorescent dyes \cite{refaat2022, tripathi2025} that are prone to photobleaching \cite{hassanpoor2021, liu2015} driven by retinal pigments and lipofuscin, thereby limiting their use in long-term or repeated measurements. Therapeutic approaches for conditions such as retinal neovascularization or intraocular tumours typically involve systemic delivery or localized injections \cite{ramsay2023, wang2019, haase2024, lee2025}, which can require high dosages and may lead to off-target effects \cite{alnamaeh2024, omoti2006}. These limitations (Figure 3A) have motivated the development of optical probes that combine high photostability with the potential for targeted delivery.

\begin{figure}[htbp]
 \centering
        \includegraphics[width=1\textwidth]{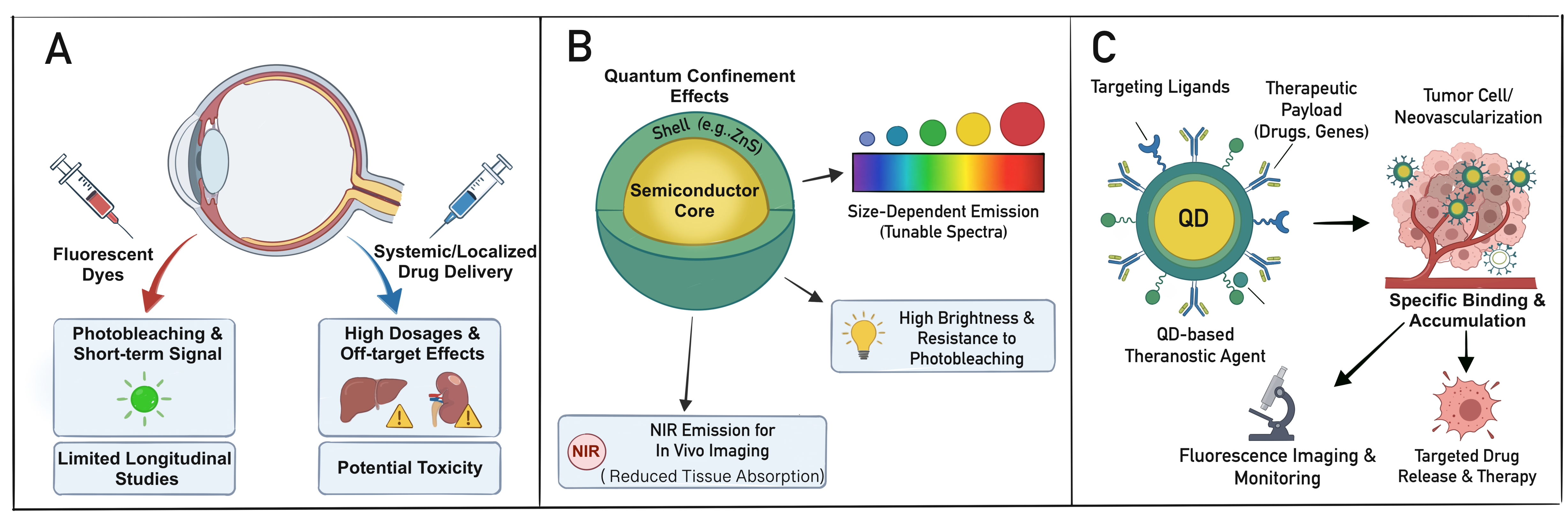}
 \caption{Schematic overview of quantum dots (QDs) for ophthalmic imaging and targeted delivery.  A. Limitations of current clinical approaches, including the rapid photobleaching of conventional fluorescent dyes and the off-target effects associated with systemic drug delivery \cite{refaat2022, tripathi2025, hassanpoor2021, liu2015, ramsay2023, wang2019, haase2024, lee2025, alnamaeh2024, omoti2006}.  B. The structural and optical advantages of QDs, highlighting quantum confinement effects, size-tunable emission spectra, and superior photostability \cite{singh2020, noh2025, lee2022, bailes2020}.  C. The design of QD-based theranostic platforms, illustrating surface functionalization with targeting ligands and the conjugation of therapeutic payloads for targeted tumour accumulation and real-time fluorescence monitoring \cite{ghafary2017, xie2025, othman2024, shen2025, diab2025, probst2013, badilli2020, qi2021}.}
\label{fig3}
\end{figure}

\subsection{Quantum approaches}
Quantum dots (QDs) are semiconductor nanoparticles whose optical properties arise from quantum confinement effects, leading to size-dependent emission spectra and high fluorescence efficiency \cite{singh2020} (Figure 3B). They exhibit tunable emission, high brightness, and improved photostability compared to conventional dyes \cite{noh2025, lee2022, bailes2020}, making them attractive candidates for biomedical imaging \cite{ghafary2017}. While distinct from quantum optical measurement approaches, QDs provide a nanoscale platform in which quantum mechanical effects directly determine optical response. Beyond their role as exogenous contrast agents, semiconductor quantum dots are also emerging as deterministic single-photon sources \cite{Hauser2026Deterministic}. Advances in InAs/InP quantum dots have enabled the generation of photons with coherence times exceeding the Fourier limit at telecom C-band wavelengths \cite{Wells2023Coherent}. The development of such coherent sources is particularly relevant for ophthalmology, as telecom wavelengths lie within the shortwave infrared (SWIR) window, which offers reduced scattering and enables deeper photon penetration in tissue \cite{roblyer2025}

QDs can be engineered to emit in spectral regions with reduced tissue absorption, including the near-infrared, facilitating in vivo imaging and labelling of cells and tissues \cite{li2025, walling2009}. Their high quantum yield and resistance to photobleaching make them well-suited for longitudinal imaging studies, where signal stability over time is essential \cite{le2023}. These properties have motivated their use in experimental ophthalmic imaging for enhanced contrast and sensitivity.

\subsection{Drug delivery and theranostics}
Surface functionalization enables QDs to bind selectively to specific molecular or cellular targets, allowing their use as targeted imaging agents \cite{ghafary2017, xie2025, othman2024} (Figure 3C). In addition, QDs have been investigated as carriers for therapeutic payloads, including drugs and gene delivery vectors, with the aim of improving localization and reducing systemic exposure \cite{probst2013, badilli2020}. These combined capabilities have led to the concept of QD-based theranostic platforms, in which imaging and therapeutic functions are integrated within a single nanoscale system.

Preclinical studies have demonstrated the feasibility of such approaches. For example, engineered QDs have been used to achieve targeted accumulation in tumour cells and to monitor therapeutic response through fluorescence imaging \cite{qi2021}. While these results highlight the potential of QDs for combined imaging and intervention, most demonstrations remain at the experimental or animal-model stage.

\subsection{Additional applications and limitations}
Beyond imaging and delivery, QDs have also been explored as photosensitizers in photodynamic therapy, where light-induced generation of reactive oxygen species can be used to damage targeted cells \cite{uprety2022semiconductor}. QD-based biosensing platforms have further been developed to detect disease-related biomarkers through fluorescence-based assays \cite{ding2022recent, mousavi2022pivotal}.

Despite these promising properties, significant challenges remain. Concerns regarding long-term toxicity, biocompatibility, and clearance, particularly for heavy-metal-based QDs, pose major barriers to clinical translation \cite{ju2026quantum}. In addition, reproducible large-scale synthesis and regulatory approval remain unresolved. As a result, while QDs offer a versatile platform for imaging and targeted delivery, their integration into routine ophthalmic practice will require further advances in materials design and safety validation.

\section{Quantum-limited vision and structured light perception}
\subsection{Current status}
The human visual system operates near the physical limit of sensitivity, with rod photoreceptors capable of responding to individual photons under appropriate conditions \cite{pugh2018}. This extreme sensitivity has motivated efforts to probe visual perception at the quantum limit using controlled light sources. At the same time, assessment of retinal structure and function typically relies on specialized imaging instrumentation, and simple visual stimuli do not directly access many underlying optical and structural properties of the retina. As a result, there is growing interest in approaches that use well-controlled optical fields to probe visual function under defined physical conditions.

\subsection{Single photon technologies in ophthalmic imaging and sensing}
Advances in single-photon detection technologies, including avalanche photodiodes, SPADs, and superconducting nanowire detectors (SNSPDs), enable measurements under extremely low light levels with high temporal resolution and low noise \cite{cova1996avalanche, stipcevic2010characterization, gao2025pixels, bian2023high, wang2025high, crockett2022high, chen2026low}. In ophthalmic contexts, these detectors allow imaging and sensing in photon-limited regimes by recording the spatial and temporal statistics of detected photons \cite{holmes2017measuring, rieke1998single}. When combined with techniques such as adaptive optics, single-photon detection can support high-resolution retinal imaging under low-light conditions and enable functional measurements such as fluorescence lifetime imaging \cite{liu2026adaptive}.

These approaches also provide a platform for studying visual perception at the single-photon level. Tinsley et al. (2016) demonstrated that human observers can detect individual photons with a probability significantly above chance \cite{tinsley2016direct}, establishing a direct link between quantum optical measurements and visual perception. At the same time, practical implementation remains challenging: single-photon systems require precise calibration, generate high data rates, and must be carefully adapted to meet safety and usability requirements in ophthalmic settings.

\subsection{Human vision as a biological detector for quantum states}
Although the human eye cannot directly perceive entanglement as a visual property, it can respond to individual photons \cite{tinsley2016direct} and therefore provides an unusual biological platform for testing detection at the quantum limit. In this context, the relevant question is not whether entanglement is consciously perceived, but whether the visual system can act as a noisy threshold detector for few-photon quantum states whose correlations are verified statistically over many trials.

Recent theoretical and psychophysical studies \cite{gassab2024conditions, vivoli2016what} have explored this question using entangled photon states and models of visual response that include photon loss, biological noise, and false-positive events. Such work treats the eye as part of a measurement chain and asks whether signatures of bipartite or multipartite entanglement could remain distinguishable after transmission through the visual system. Entanglement witnesses and related statistical tools provide a framework for assessing these limits, while also clarifying the distinction between single-photon sensitivity and access to nonclassical correlations.

\subsection{Perception of orbital angular momentum and nonseparable states}
Orbital angular momentum (OAM) of light \cite{Yao2011}, characterized by an azimuthal phase structure $\exp(i\ell\theta)$, forms an unbounded basis of spatial eigenmodes. Unlike polarization, which spans a two-dimensional Hilbert space \cite{Subramanian2017}, OAM modes are theoretically infinite-dimensional \cite{Fang2021, Dorrah2018}. Direct perception of individual OAM eigenvalues is inaccessible to the human eye, \cite{Dorrah2018} as it lacks the modal decomposition capabilities required to resolve such states. However, classical light fields that encode OAM through structured intensity and polarization profiles can evoke distinct percepts in the human visual system, particularly when combined with retinal structures that support polarization-dependent scattering \cite{pushin2023structured, pushin2024psychophysical, Sarenac2020, Sarenac2022}.

Recent psychophysical experiments have demonstrated that spin-orbit structured light can enhance classical entoptic phenomena such as Haidinger’s \cite{pushin2025characterizing, pushin2023structured, pushin2024psychophysical, sarenac2024structured, sarenac2025roadmap} and Boehm’s brushes \cite{pushin2025topological, sarenac2025roadmap}. In the foveal region, Haidinger’s brush arises from dichroic absorption of polarized light by the macular pigment \cite{mottes2022haidinger} in the radially organized Henle fibre layer \cite{ramtohul2023oct}. In the periphery, Boehm’s brush is believed to result from polarization-dependent scattering by obliquely oriented nerve fibres \cite{pushin2025topological, temple2015perceiving}. Both phenomena exhibit sensitivity to vector vortex beams, where the spatial and polarization degrees-of-freedom are nonseparable \cite{pushin2025topological}. These beams form classical analogues to quantum entangled states and can be modelled as nonseparable superpositions across multiple degrees-of-freedom \cite{mclaren2015measuring}.

Experimental stimuli combining circular polarization and azimuthally varying OAM modes have been used to evoke entoptic patterns (Figure 4) with structured azimuthal lobes. Participants could identify the rotation direction of these patterns and their threshold eccentricity, revealing the perceptual integration of nonseparable spin-orbit content \cite{kapahi2024measuring}. The human eye thus functions as a spatial integrator that is sensitive to the classical signatures of OAM when encoded in polarization-modulated intensity fields \cite{pushin2025characterizing}.

\begin{figure}[htbp]
 \centering
        \includegraphics[width=1\textwidth]{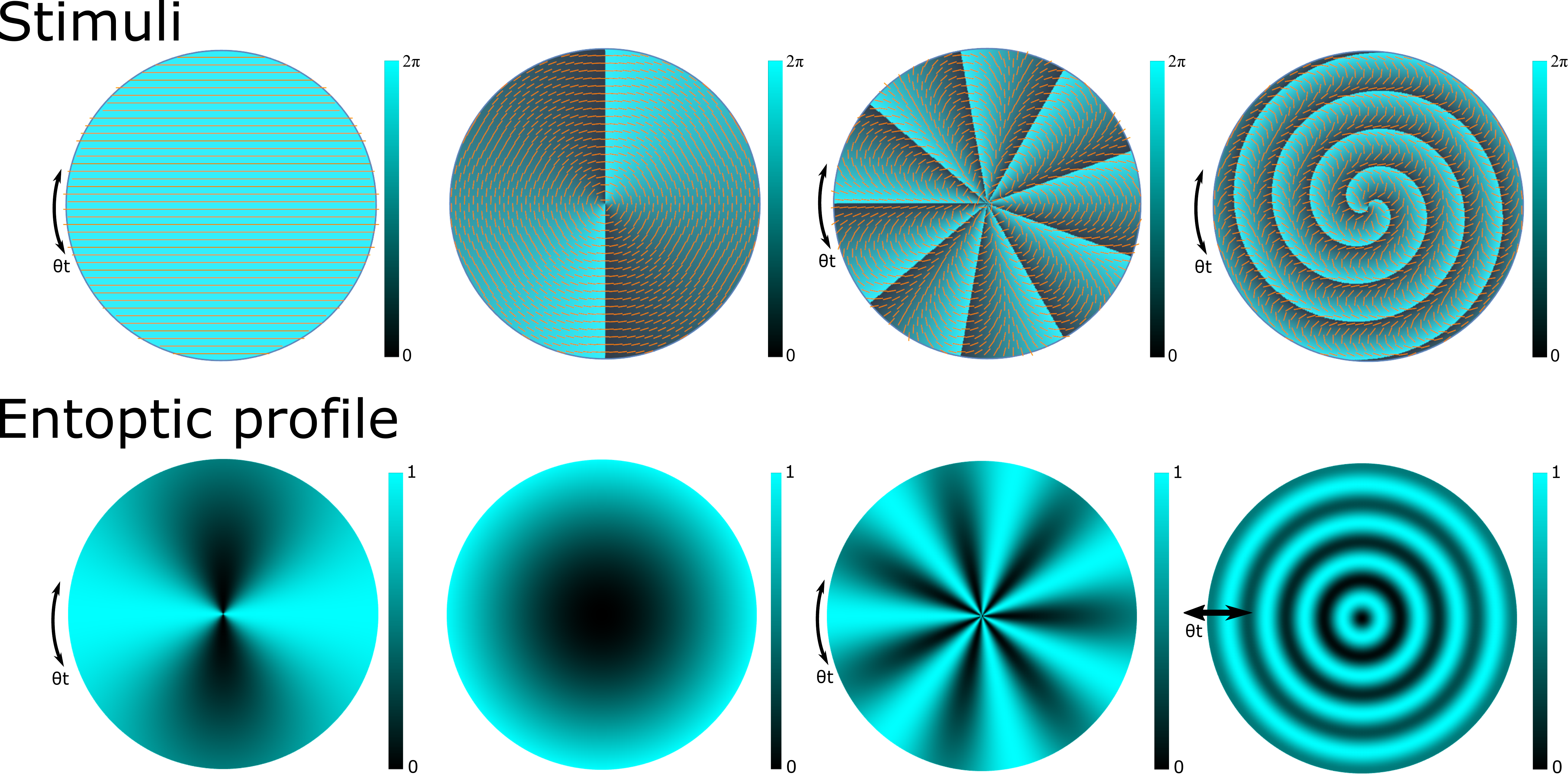}
 \caption{Examples of the phase and polarization patterns of structured light stimuli (top row) alongside the corresponding entoptic images perceived by an individual with a healthy macula (bottom row). The sharpness of these entoptic images is directly related to the density of macular pigment, which is highest at the fovea and diminishes toward the periphery. In the first column, a stimulus with horizontal polarization is shown, resulting in the classic Haidinger’s brush effect. By employing structured light techniques to generate stimuli with polarization-coupled orbital angular momentum (OAM) states, a diverse range of entoptic patterns can be produced. The second column demonstrates the use of a stimulus with OAM = 2, which aligns with the structure of the Henle fibres and produces a uniform entoptic pattern. The third column shows that increasing the OAM value leads to stimuli that generate more pronounced entoptic effects, characterized by a greater number of azimuthal fringes. The final column presents a stimulus combining a radial state with OAM = 2, resulting in entoptic profiles featuring radially varying fringes.}
\label{fig4}
\end{figure}

This class of stimuli opens opportunities for probing the perceptual accessibility of nonseparability using quantum-inspired tools. Future experiments could leverage advanced quantum-like structured beams such as ray-wave \cite{shen2020structured, he2022towards} or SU(2) coherent-state modes \cite{shen2021creation, wu2025off}, which support polarization-OAM-trajectory coupling and allow tunable control of perceived spatial structure. In such experiments, entanglement witnesses adapted from quantum optics (such as Bell-like inequalities \cite{liang2015family} or state tomography protocols \cite{zhu2010minimal}) may be applied to characterize the strength of classical nonseparability in stimuli that evoke visual percept. Structured entoptic responses to these beams could serve as sensitive probes of retinal birefringence, macular pigment distribution, and peripheral visual function, motivating further development of quantum-inspired diagnostics in ophthalmology.

\section{Conclusion}
This review has examined the emerging role of quantum and quantum-inspired approaches in ophthalmology, with a focus on imaging, sensing, and the probing of visual function. Across these areas, a common theme is the operation of both instrumentation and the visual system itself under photon-limited conditions, where the fundamental properties of light and detection become critical.

Q-OCT and related interferometric approaches illustrate how nonclassical correlations can be used to modify measurement strategies, offering advantages such as dispersion cancellation \cite{Abouraddy2002} and improved performance in low-light regimes. In parallel, advances in single-photon detection enable imaging and sensing at extremely low photon flux, while also providing a framework for studying visual perception at the level of individual quanta. Correlation-based imaging techniques, including ghost imaging, further demonstrate alternative approaches to image formation that do not rely on conventional spatial detection, although their practical application remains limited by efficiency and acquisition speed.

At the same time, nanoscale systems such as quantum dots provide complementary capabilities for biomedical imaging and targeted delivery. While their optical properties are governed by quantum confinement rather than nonclassical light, they represent an important class of probes for enhancing contrast and enabling molecular specificity. However, challenges related to toxicity, biocompatibility, and clinical translation remain significant.

Beyond instrumentation, structured light and controlled optical fields offer new ways to probe the visual system itself. Experiments at the single-photon level and with spin-orbit structured beams highlight both the sensitivity and the limitations of human vision as a detector, and provide insight into how spatial and polarization information is processed by the retina.

Despite substantial progress, most quantum-enabled approaches remain in the early stages of development. Key challenges include system complexity, detector performance, integration with existing clinical platforms, and the need for rigorous validation under realistic imaging conditions. Future advances will likely depend on combining quantum and classical techniques, rather than on replacing established methods.

Taken together, these developments suggest that quantum and quantum-inspired approaches will play an increasing role in extending the capabilities of ophthalmic imaging and in providing new tools for studying visual function. This steady integration of quantum principles promises to fundamentally redefine the precision limits of ophthalmic diagnostics and therapeutics, paving the way for a new era of vision science.

\section*{Acknowledgments}
This study was supported by the InnoHK initiative of the Innovation and Technology Commission of the Hong Kong Special Administrative Region Government, the Canada First Research Excellence Fund (CFREF), the New Frontiers in Research Fund (NFRF), the Canadian Excellence Research Chairs (CERC) programme and the Natural Sciences and Engineering Research Council of Canada (NSERC). We acknowledge Raushan Toktarova for preparing the illustrations in Figure 3.

\bibliographystyle{vancouver}
\bibliography{references}

\end{document}